\documentstyle[12pt, epsf]{article}
\input math_macros.tex

\def\ref#1{$^{#1)}$}

\def\sl{\slash}
\begin{document}
\begin{titlepage}
\begin{center}
\today     \hfill    LBL-35973 \\
%          \hfill    UCB-PTH-xx/xx \\

\vskip .1in

{\large \bf Quantum Electrodynamics at Large Distances III:
Verification of Pole Factorization and the Correspondence Principle}
\footnote{This work was supported by the Director, Office of Energy
Research, Office of High Energy and Nuclear Physics, Division of High
Energy Physics of the U.S. Department of Energy under Contract
DE-AC03-76SF00098, and by the Japanese Ministry of Education,
 Science and Culture under a
Grant-in-Aid for Scientific Research (International Scientific Research Program
03044078).}

\vskip .1in
Takahiro Kawai\\

{\em Research Institute for Mathematical Sciences\\
Kyoto University\\
Kyoto 606-01 JAPAN\\}
\vskip .1in
Henry P. Stapp \\

{\em     Lawrence Berkeley Laboratory\\
    University of California\\
   Berkeley, California 94720}

\end{center}

\vskip .1in

\begin{abstract}

In two companion papers it was shown how to separate out from a scattering
function in quantum electrodynamics a distinguished part that meets the
correspondence-principle and pole-factorization requirements.
The integrals that define the terms of the remainder are here shown to have
singularities on the pertinent Landau singularity surface that are weaker than
those of the distinguished part. These remainder terms therefore vanish,
relative to the distinguished term, in the appropriate macroscopic limits.
This shows, in each order of the perturbative expansion, that quantum
electrodynamics does indeed satisfy the pole-factorization and
correspondence-principle requirements in the case treated
here. It also demonstrates the efficacy of the computational techniques
developed here to calculate the consequences of the principles of quantum
electrodynamics in the macroscopic and mesoscopic regimes.
\end{abstract}
\end{titlepage}
%THIS PAGE (PAGE ii) CONTAINS THE LBL DISCLAIMER
%TEXT SHOULD BEGIN ON NEXT PAGE (PAGE 1)
\renewcommand{\thepage}{\roman{page}}
\setcounter{page}{2}
\mbox{ }

\vskip 1in

\begin{center}
{\bf Disclaimer}
\end{center}

\vskip .2in

\begin{scriptsize}
\begin{quotation}
This document was prepared as an account for work sponsored by the United
States Government.  Neither the United States Government nor any agency
thereof, nor The Regents of the University of California, nor any of their
employees, makes any warranty, express or implied, or assumes any legal
liability or responsibility for the accuracy, completeness, or usefulness
of any information, apparatus, product, or process disclosed, or represents
that its use would not infringe privately owned rights.  Reference herein
to any specific commercial products process, or service by its trade name,
trademark, manufacturer, or otherwise, does not necessarily constitute or
imply its endorsement, recommendation, or favoring by the United States
Government or any agency thereof, or The Regents of the University of
California.  The views and opinions of authors expressed herein do not
necessarily state or reflect those of the United States Government or any
agency thereof of The Regents of the University of California and shall
not be used for advertising or product endorsement purposes.
\end{quotation}
\end{scriptsize}

\vskip 2in

\begin{center}
\begin{small}
{\it Lawrence Berkeley Laboratory is an equal opportunity employer.}
\end{small}
\end{center}

\newpage
\renewcommand{\thepage}{\arabic{page}}
\setcounter{page}{1}
%THIS IS PAGE 1 (INSERT TEXT OF REPORT HERE)

\noindent {\bf 1. Introduction}
\vskip 9pt

In papers I$^1$ and II$^2$ we examined the functions $F(g)$ associated with
the infinite set
of graphs $\{g\}$ obtained by dressing a simple triangle graph $G$ with soft
photons in all possible ways. A distinguished set of contribution to these
functions $F(g)$ was singled out and called ``dominant'' because these
contributions were expected to dominate the macroscopic behaviour of the
scattering functions. Each of these distinguished parts was shown to be well
defined, and to have a singularity of the form $\log \varphi$ on the
(Landau-Nakanishi) triangle-diagram singularity surface $\varphi = 0$.
This form $\log \varphi$ agrees with the form of the singularity of the
original Feynman function $F(G)$ on $\varphi = 0$, and it produces the same
kind of large-distance fall-off. Moreover, these dominant contributions yield
(exactly once) every term in the perturbative expansion of the triangle-diagram
version of the pole-factorization property
$$
\mbox{Disc} F'|_{\varphi =0}=F'_1F'_2F'_3.
$$
The left-hand side of this equation represents value at $\varphi =0$ of the
discontinuity across the surface $\varphi = 0 $ of the function $F'$ that is
obtained by omitting the contributions from the ``classical'' photons. These
latter contributions are supplied by the unitary operator $U(L)$ --- after the
transformation to coordinate space. Consequently, the  discontinuity formula
given above entails that the contributions from these ``dominant'' terms give
just the classical-type large-distance behaviour demanded by the
correspondence principle: the rate of fall-off at large distances is exactly
what follows from the classical concept of three stable charged particles,
each moving from one scattering region to another, and the electromagnetic
field generated by $U(L)$ is exactly the quantum analog of the classical
electromagnetic field generated by the motions of these three charged
particles. In the present article we shall show that, in each order of the
perturbation expansion, the terms of the remainder  give {\it no} contributions
to the discontinuity defined above. Consequently, these ``non-dominant'' terms
give no contribution to the leading term in the the asymptotic large-distance
behaviour, and hence the correspondence-principle requirement is satisfied.

In section 2 we examine the simplest example, namely the triangle graph $G$
dressed with {\it one} internal soft photon. The remainder part is separated
into a sum of terms. For some terms the weakening of the singularity on
the surface $\varphi =0$ is  associated with the topological complexity of the
graph that represents this term, namely its non-separability: cutting the
graph at the three $*$ lines associated with the three Feynman-denominator
poles does not separate the graph into three disjoint parts. This means that
the integration over the momenta of the internal photons tends to shift the
position of the singularity, and hence weaken it. For the remaining terms the
weakening of the singularity on $\varphi =0$ is due to the replacement of one
or more of the three pole singularities $(p_s^2-m^2)^{-1}$ of the integrand by
a  pair of logarithmic singularities: this replacement of the pole
singularities in the integrand by logarithmic singularities likewise leads to a
weakening of the singularity of the integral on $\varphi=0$.

Our problems here are first to show that these reductions in the degree of the
singularity on $\varphi =0$, which emerge easily within our formalism
in this simple one-photon example, hold for every $g$ obtained by dressing
the triangle graph $G$ with soft photons,
and second to show that the weakening of the singularity is,
in every case, a weakening by at least one full power of $\varphi$, up to
a prescribed finite number of powers of $\log \varphi$ that increases
linearly with the number of photons, and hence with powers of the coupling
constant. This strong result means that the validity of the
correspondence--principle in the
large--scale limit, which is established here at each order of the perturbative
expansion, cannot be upset by an accumulation of powers of
$(\log \varphi)^n$ that leads to a singularity of the form
$\varphi^{-\beta}$, where $\beta$ is of the order of the fine-structure
constant
$(\sim 1/137)$. Accumulations of this kind occur often in field theories.
The appearance of the fine-structure constant in the exponent arises from the
fact that usually, just as in our case, the number of powers of
$\log \varphi$ is linearly tied to the number of powers
of the fine-structure constant. We have not studied, in our case, the numerical
factors that multiply the terms of the remainder, except to show that they
are all finite. Hence we can make here no claim pertaining to meaningfulness of
the infinite sum in our case: we plan to examine this question later.

To establish our general conclusions we need two auxiliarly results.
The first is a geometric property concerning the structure of the
Landau-Nakanishi surface. It is proved in section 3. The second
pertains to several singular integrals. The needed computation is performed
in section 4. The required properties of the various intergrals are then
proved in sections 5, 6, and 7.
\newpage
\noindent {\bf 2. Examination of non-dominant singularities for the
one-photon case.}

\vskip 9pt

\begin{figure}
\caption
{ A graph $g$ representing a soft-photon correction to a hard-photon
triangle-diagram process $G$. The letters Q near the ends of the
wiggly line that represents the soft photon indicate that this particle is
coupled to the charged particle through the ``quantum'' part of the full
quantum-electrodynamical coupling.}
  \epsfxsize = 4.52in
  \epsfysize = 5.89in
  \epsffile{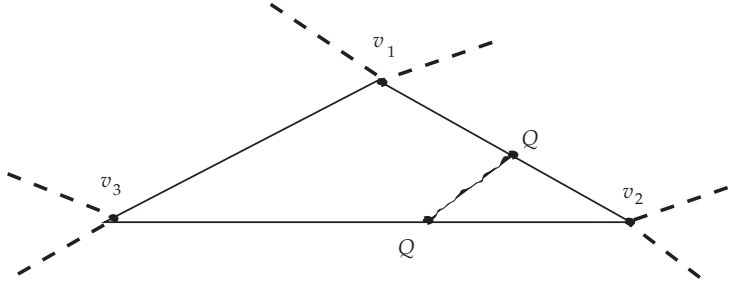}
  \label{fig1}
\end{figure}

Let us consider the contributions associated with the graph $g$
shown in Figure \ref{fig1}.

\begin{figure}
\caption
{ Four $*$ graphs representing the four terms in the meromorphic part
of the function represented by the graph in fig. 1. These four terms arise from
a decomposition of the meromorphic parts associated with each of the three
sides of the triangle into poles times residues. The $*$ lines represent
Feynman-denominator poles. The other charged-particle lines represent residue
factors.}
\epsfxsize = 6.35in
  \epsfysize = 4.53in
  \epsffile{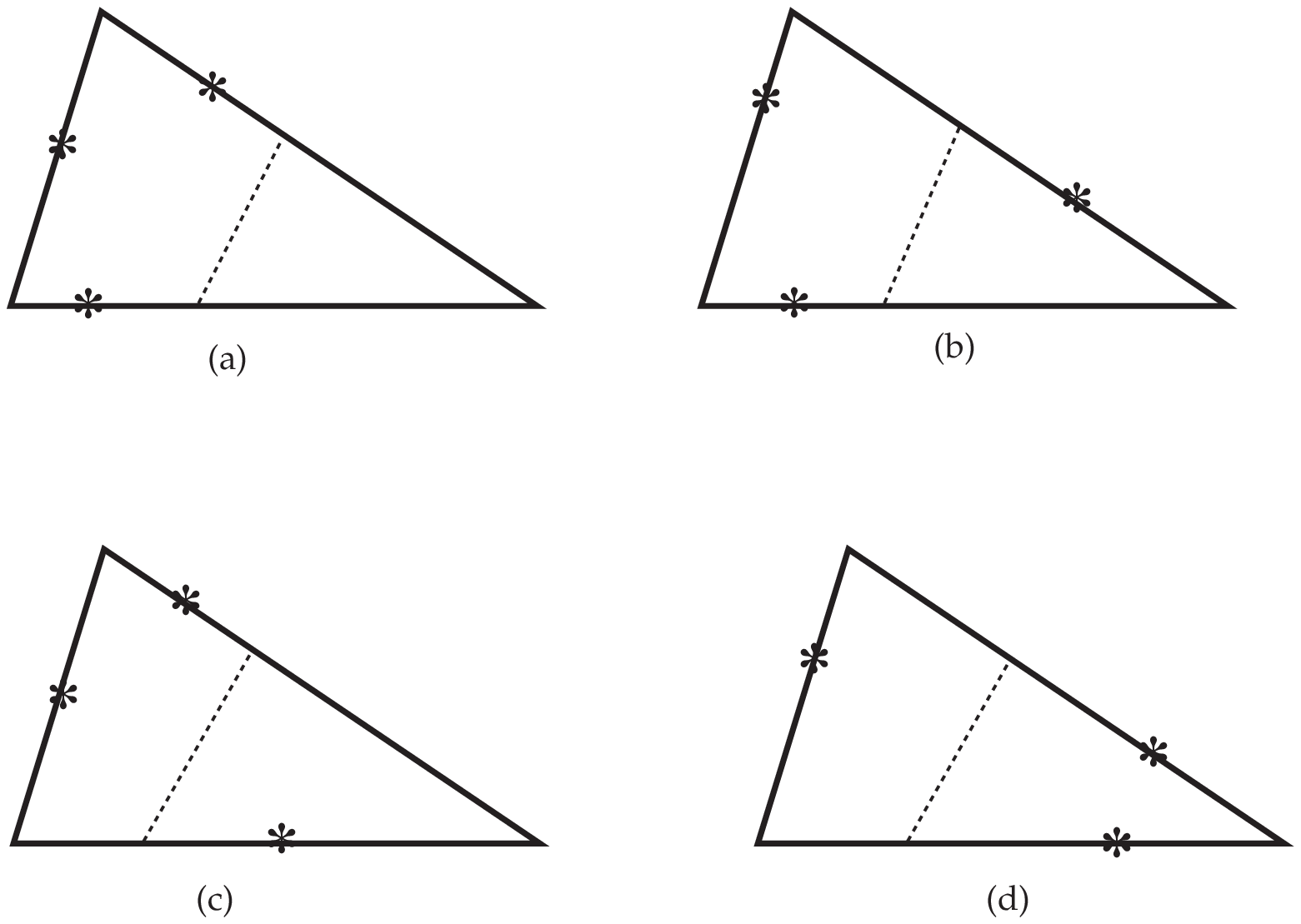}
  \label{fig2}
\end{figure}

In references 1 and 2 we showed how to separate the contribution
represented by the graph of Fig. 1 into a
meromorphic part consisting of a sum of the four terms represented by the
the four $*$ graphs of Figure \ref{fig2} , plus a ``non-meromorphic'' part.

\noindent The term associated with the graph $a$ of Figure 2 is separable, and
is classified
as dominant.  The associated function $F_a$  is given by (4.1) and (4.3) of
Ref. 1, and has a logarithmic singularity along the Landau surface $\varphi
=0$.
We also claimed there that the other three  terms in the meromorphic part are
are infrared finite, and have weaker singularities along the
Landau surface $\varphi =0$. In the following subsection we verify this claim
for the case of the function $F_b$ associated with graph (b) of Fig. 2. The
other two cases, $c$ and $d$, can be treated similarly.
\newpage
\noindent {\bf 2.i. The contribution from a one-photon nonseparable meromorphic
 part.}

The function $F_b$ was given in (4.4) of Ref. 1:
$$
\eqalignno{
F_b &= \int {d^4 p\over (2\pi )^4}
  \int^\delta_0 2rdr \int
  {d^4\Omega\over(2\pi )^4} \ \
  {
  i\delta (\Omega^2_0 + \vec{\Omega}^2-1)\over \Omega^2 + i0
   }\cr
&\Tr \Bigg\{ {i(\sl{p}+m)\over p^2-m^2} V_1
\bigg(
  {(2p_{i\mu}+2r\Omega_\mu )
 \Omega^2(2p_1\Omega + 2 r\Omega^2)^{-1} -
\slash{\Omega} \gamma_\mu\over 2p_1 \Omega + r \Omega^2}\bigg)\cr
&\times {(\sl{p}_1 + r\sl{\Omega} + m)\over (p_1+ r\Omega )^2-m^2}
  V_2\cr
&\times \bigg(
  {
    2p_{2\mu}
    \Omega^2(2p_2\Omega)
    -\gamma_\mu \sl{\Omega}\over 2p_2\Omega + r \Omega^2
  }
   \bigg)
 {(\sl{p}_2 +m)\over p^2_2 - m}
  V_3 \bigg\}
&(2.ii.1)\cr}
$$
It was shown in Ref. 2 that we can distort the
$\Omega$-contour so that {\it Im\/}$\Omega^2 > 0$ at $\Omega^2=0$, and
{\it Im\/}$ p_j \Omega > 0\ (j=1,2)$ at $p_j\Omega=0$. Then, except for three
pole-factors $p^2-m^2, (p_1+ r\Omega)^2 - m^2$ and $p^2_2 - m^2$, each
denominator of the integrand of $F_b$ is different from
zero.

The $r$-integration $\int^\delta_0 rdr/[(p_1+r\Omega)^2 - m^2$] can be
explicitly performed, and when $p_1 \Omega \neq 0$ its dominant singularity
along $p^2_1=m^2$ is
$$
- {(p^2_1 -m^2) \log (p^2_1 - m^2)\over 4(p_1\Omega)^2}.
$$
Combining this singularity, instead of the ordinary pole $1/(p^2_1-m^2)$, with
the other two poles, i.e., $1/(p^2-m^2)$ and $1/(p^2_2-m^2)$, we perform the
$p$-integration and  find a singularity $A(q,\Omega) \varphi (q)^2 \log \varphi
(q)$, with $A$ being analytic.
Performing the $\Omega$-integration along the compact
 distorted contour, the dominant
singularity of $F_b$ is $\varphi^2 \log \varphi$.

Essentially the same argument covers the case where one of the two meromorphic
parts is due to a $C$-coupling. Then the factor $rdr$ becomes simply $dr$, and
the singularity becomes $\varphi \log \varphi$.
\newpage

\noindent{\bf 2.ii. The contribution from a pair of non-meromorphic
parts arising from  one photon.}

\medskip

The contribution of $I$ in (4.6) in Ref. 1 to the amplitude is
$$
\eqalignno{
F &= \int_{|\Omega |=1} {d^4\Omega\over \Omega^2+i0}
    \int^\delta_0 {dr\over r} \int d^4p_3
   {1\over p^2_3 - m^2 + i0} \log {(q_1+ p_3 + r\Omega)^2
   - m^2+i0\over (q_1 +p_3)^2 - m^2 +i0}\cr
& \times \log
   {(p_3 - q_3 -r\Omega)^2 - m^2 + i0\over (p_3 - q_3)^2 - m^2
+i0},&(2.ii.1)\cr}
$$
with the $p_i$ defined as in Fig. 1 of ref.2.
Here the  $\Omega$-contour is deformed so that {\it Im\/}$\Omega^2 >0$ and
{\it Im\/}$ p_j
 \Omega >0$ (j=1,2).  Performing the $p_3$-integration we find
$$
\eqalignno{
\int_{|\Omega|=1} {d^4\Omega\over \Omega^2 +i0}
& \int^\delta_0 {dr\over r} (G(q_1 + r\Omega, q_3 + r\Omega) \cr
&-
 G(q_1 + r\Omega, q_3) - G(q_1, q_3 + r\Omega) +
 G(q_1, q_2)). &(2.ii.2)\cr}
$$
where
  $$
G(q_1,q_2)=\varphi(q_1, q_2)^2 \log (\varphi (q_1, q_2) + i0).
$$
Since
$$
\eqalignno{
(\partial/\partial r) \varphi (q_1 + r\Omega, q_3 + r\Omega )& =
(\partial\varphi/\partial q_1 + \partial\varphi/\partial q_3) \cdot \Omega \cr
& = (
\alpha_1\rho_1 + \alpha_2\rho_2)\cdot \Omega = - \alpha_3 p_s \cdot \Omega \neq
0
\cr}
$$
 holds by the Landau equation, we can find non-vanishing functions $a(q_1, q_3,
r,
 \Omega)$ and $b(q_1,q_3)$ for which
$$
\varphi (q_1 +r\Omega, q_3 + r\Omega) = a(q_1, q_3, r, \Omega)(r - b(q_1, q_3)
\varphi (q_1 + r\Omega, q_3))
$$
holds (Cf.$\S$ 3 below). Similar decompositions hold also for $\varphi
(q_1 + r\Omega,q_2)$ and $\varphi (q_1, q_2 + r\Omega)$.
Hence application of the results in $\S$ 4 below to
$\int^\delta_0 {dr\over r} (G(q_1 +r\Omega, q_3 + r\Omega)
- G(q_1, q_3))$ etc.
entails that the $r$-integration in (2.ii.2) produces a singularity of the form
$\varphi (q_1, q_3)^2 (\log (\varphi (q_1, q_3) + i0)^2$ near $\varphi =0$.
Since the $\Omega$-integration (along a suitably detoured path) is over the
compact set, $F$ itself behaves as $\varphi^2(\log (\varphi + i0))^2.$

\newpage
\noindent{\bf 2.iii. The contribution from a coupling of a non-meromorphic
part with either a meromorphic part or a $C$-part.}

\medskip

If a meromorphic part is coupled with a non-meromorphic part, the RHS of
(2.ii.1) is replaced by an integral of the following form:
$$
\eqalignno{
\int_{|\Omega|=1} {d^4\Omega\over \Omega^2+i0} &\int^\delta_0 dr \int d^4p_3
{1\over p^2_3-m^2+i0} \log {(q_1+p_3+r\Omega)^2 - m^2+i0\over (q_1 + p_3)^2 -
m^2+i0}\cr
&{1\over (p_3 -q_3)^2 - m^2+i0} \ \ {1\over 2(p_3-q_3)\Omega + r\Omega^2+i0}.
&(2.iii.1)\cr}
$$
By deforming the $\Omega$-contour, in the manner specified in ref. 2,
so that {\it Im\/}$\Omega^2 >0$
and {\it Im\/}$ p_j \Omega >0 (j=1, 2)$ [with $p_1=q_1+p_3, p_2=p_3 - q_3$],
we find the singularity of this integral near
$\varphi (q_1, q_3)=0$ is $\varphi \log (\varphi +i0)$, as there is no
potentially divergent factor $1/r$.

If the meromorphic part is replaced by a $C$-term, then the dominant
singularity is given by an integral similar to (2.iii.1) but with the
replacement of
the residue factor $1/(2(p_3-q_3)\Omega + r\Omega^2)$ by $1/r(p_3-q_3)\Omega$.
Hence a potentially divergent factor $1/r$ arises.
But this problem is circumvented by combining the singularity originating from
$\log ((q_1 +p_3 +r\Omega)^2 - m^2 +i0)$ and that from $\log ((q_1+p_3)
^2-m^2+i0)$;
the results in $\S$ 4 show, with a reasoning similar to (but simpler than)
that
in $\S$ 2.ii, that the resulting singularity is $\varphi (\log (\varphi + i0))
^2$.

\newpage
\noindent{\bf 3. A normalization of the function defining a
Landau surface.}
\medskip

The purpose of this section is to prove the following lemma, which is an
adaptation of the implicit function theorem (or the Weierstrass preparation
theorem in the theory of holomorphic functions of several variables) to the
Landau surface shifted by a vector $\Delta$ determined by photons
that bridge star lines. (cf.$\S$ 11. of Ref. 1.).
Here and in what follows, $(r_1, ..., r_n)$ denotes a nested set of polar
coordinates introduced in Ref. 1, $\S$5.

\noindent{\bf Lemma 3.1}
Let $\varphi (q)$ denote a defining function of the Landau surface for the
triangle diagram and let $\hat{q}$ be a point on the surface.
Let $i$ be the smallest $j$ such that $j$ identifies a  bridge photon line.
(A {\it bridge} photon line is a photon line that has meromorphic couplings on
both ends and that completes --- via the rules defined below Eqn. (2)of ref. 2
 --- to a
closed photon loop that  passes along at least one star line.)
Then on a sufficiently small neighborhood of $q_0$ and for sufficiently small
$\rho_i = r_1 \cdots r_i$ there exist non-vanishing holomorphic functions
$B(q, \rho_i,
k'/\rho_i)$ and $C(q, k'/\rho_i)$ such that
$$
\varphi (q-\Delta) = B(q, \rho_i, k'/\rho_i) (\rho_i - \varphi
(q)/C(q,k'/\rho_i))
\eqno(3.1)
$$
holds, where $k'$ denotes the collection of bridge lines.

\noindent{\bf{Proof.}} Since $i$ is the first bridge photon line, any bridge
photon line $k_\ell$ has the form $k_\ell= \rho_i r_{i+1}\cdots r_\ell
\Omega_\ell$.
Hence $k'/\rho_i$ is actually independent of $\rho_i$.
Furthermore, as is shown at the beginning of section 5 ((5.1)),$\partial
\varphi (q-\Delta)
/\partial \rho_i|_{\rho_i=0}\neq 0$ holds.
Hence the Weierstrass preparation theorem guarantees the local and unique
existence of a non-vanishing holomorphic function $B(q, \rho_i, k'/\rho_i)$,
and a
holomorphic function $R(q, k'/\rho_i)$, which vanishes for $q_i=q_0$,
for which the following holds:
$$
\varphi (q-\Delta)= B(q, \rho_i, k'/\rho_i) (\rho_i - R(q,
k'/\rho_i))\eqno(3.2)
$$
Setting $\rho_i = 0$ in (3.2) we find
$$
\varphi (q) = B(q, 0, k'/\rho_i) (-R(q, k'/\rho_i),
$$
that is,
$$
R(q, k'/\rho_i)=\varphi (q)/(-B(q,0, k'/\rho_i)).
$$
Hence by choosing $C(q, k'/\rho_i)=-B(q, 0, k'/\rho_i)$ we obtain (3.1).

\newpage

\noindent{\bf 4. Some auxiliary integrals.
}
\medskip

The purpose of this section is to find an explicit form of the singularities of
several integrals that we encounter in dealing with infrared problems.
 The simplest
example of this sort is the following integral $I(t)$:
$$
I(t) = \int^\delta_0 [\log (r + t + i0) - \log (t+ i0)] dr/r, \ \ (\delta >0).
$$
In spite of the divergence factor $1/r$, $I(t)$ is well defined as a (hyper)
function of $t$.
In order to see this, it suffices to decompose $I(t)$ as
$$
\int^{t/2}_0 (\log (r +t) - \log t) dr/r + \int^\delta_{t/2} (\log (r+t) - \log
t) dr/r
$$
with $Im \ t > 0$: the well-definedness of the second integral is clear, while
the
fact that
$$
\log (r + t) - \log t = \log (1 +{r\over t}) \sim {r\over t}
$$
 holds in the
domain of integration of the first integral entails its well-definedness.
Furthermore $I(t)$ (thus seen to be well-defined) satisfies the following
ordinary
differential equation:
$$
t{d\over dt} I(t) =  \log (t + i0) - \log (t + \delta + i0).\eqno(4.1)
$$
Hence $(t {d\over dt})^2$ is holomorphic near $t=0$.
Then it follows from the general theory of ordinary differential equations that
$I(t)$ has the form
$$
C_2(\log (t +i0))^2 + C_1 (\log (t+i0)) + h(t),\eqno(4.2)
$$
where $C_1$ and $C_2$ are constants and $h(t)$ is holomorphic near $t=0$.
Furthermore, by substituting (4.2) into (4.1)
and comparing the coefficients of singular terms at $t=0$, we find $C_2= 1/2$.

This computation can be generalized as follows:

\medskip

\noindent{\bf Proposition 4.1.} Let $J(\alpha, j; t)\ (\alpha \neq 0, 1,2,
\cdots
; j \geq 1)$ denote the following integral:
$$
\int^{\delta=\rho_0}_0 {d\rho_1\over \rho_1} \int^{\rho_1}_0 {d\rho_2\over
\rho_2} \cdots
 \int^{\rho_{j-2}}_0
{d\rho_{j-1}\over \rho_{j-1}} \int^{\rho_{j-1}}_0 (t + \rho_j + i0)^\alpha
d\rho_j.
$$
Then the singularity of $J(\alpha, j; t)$ near $t=0$ is of the following form
with some constants $C_\ell\ (\ell =0, \cdots , j-1)$:
$$
\Bigg\{ \matrix{(t+i0)^{\alpha +1} (\sum^{j-1}_{\ell =0} C_\ell (\log (t+i0))
^\ell) , &\hbox{if} \ \alpha \neq -1\cr
          \sum^{j-1}_{\ell =0} C_\ell (\log (t + i0))^{\ell + 1}, &\hbox{if} \
           \alpha =- 1}. \eqno(4.3)
$$

\medskip

\noindent{\bf Remark 4.1.} If $\alpha$ is a non-negative integer, the integral
$J(\alpha, j; t)$ is not singular at $t=0$.

\medskip

\noindent{\bf Proof of Proposition 4.1.} The well-definedness can be verified
by the same method as was used for the above example $I(t)$.
To find its singularity structure, we again make use of an ordinary
differential
equation as follows:
$$
\eqalignno{
&(t {d\over dt} - (\alpha + 1)) J (\alpha, j; t)
    = \left( t {d\over dt} - (\alpha + 1)\right)
        \int^\delta_0 {d\rho_1\over \rho_1} \int^{\rho_1}_0 {d\rho_2\over
\rho_2}\cdots\cr
    &\int^{\rho_{j-2}}_0{d\rho_{j-1}\over \rho_{j-1}}
      ( (t+ \rho_{j-1} +i0)^{\alpha +1}
      - (t+i0)^{\alpha +1})/(\alpha +1)\cr
   &= \int^\delta_0 {d\rho_1\over \rho_1} \int^{\rho_1}_0 {d\rho_2\over \rho_2}
\cdots
   \int_0^{\rho_{j-2}} {d\rho_{j-1}\over \rho_{j-1}}
  ( t(t+\rho_{j-1} + i0)^\alpha -
   (t+\rho_{j-1} + i0 )^{\alpha +1} )\cr
  &=-\int^\delta_0 {d\rho_1\over \rho_1} \int^{\rho_1}_0 {d\rho_2\over \rho_2}
\cdots
\int^{\rho_{j-2}}_0 (t+\rho_{j-1}+i0)^\alpha d\rho_{j-1} =- J(\alpha, j-1; t)
(j\geq 2).
\cr}
$$
 Repeating this computation, we finally obtain
$$
(t {d\over dt} - (\alpha + 1))^{j-1} J(\alpha, j; t)
= (-1)^{j-1} \int^\delta_0
(t + \rho_1)^\alpha d\rho_1,
$$
and hence we find
$(t {d\over dt} - (\alpha + 1))^j J(\alpha, j; t)$
is holomorphic near $t=0$.
Again, by using the general theory of ordinary differential equations,
 we obtain
the required formula (4.3).

\medskip

\noindent{\bf Remark 4.2.} Although we do not need the exact values of
$C_\ell$'s, we note that $C_{j-1}$ in (4.3) is simply given by
$(-1)^{j-2}/(j-1)
!(\alpha +1)$ if $\alpha \neq -1$.
In order to find this value it suffices to insert (4.3) into the recurrence
relation $(t d/dt - (\alpha + 1)) J(\alpha, j; t) =- J(\alpha, j-1;t)$ and use
the trivial relation
$$
\eqalignno{
\left(t {d\over dt} - (\alpha + 1)\right) J(\alpha, 2; t) &=- \int^\delta_0 (t+
\rho_1 + i0)
^{\alpha} d\rho_1\cr
&= ((t+i0)^{\alpha+1}/(\alpha+1)) - ((t+\delta+i0)^{\alpha +1}/(\alpha +
1))\cr}
$$
as the starting point of the induction.
Similarly $C_{j-1}$ for $\alpha =- 1$ is equal to $(-1)^j/j!$.
The coefficient of the most singular term can be similarly computed explicitly
for the integrals to be dealt with in subsequent propositions.

The following modification of Proposition 4.1 is often effective in actual
computations.

\medskip

\noindent{\bf Proposition 4.2.} (i) Let $K(\alpha, j; t)\ (\alpha \neq 0, 1,
\cdots
; j \geq 1)$ denote the following integral:
$$
\int^\delta_0 {d\rho_1\over \rho_1} \int^{\rho_1}_0
 {d\rho_2\over \rho_2} \cdots \int^{\rho_j-2}_0
{d\rho_{j-1}\over \rho_{j-1}} \int^{\rho_{j-1}}_0 \rho_j (t+\rho_j + i0)^\alpha
d\rho_j.
$$
Then its singularity near $t=0$ is of the following form for some constants
$C_\ell (\ell =0, \cdots, j-1)$:
$$
\Bigg\{ \matrix{ (t+i0)^{\alpha +2} \left( \sum^{j-1}_{\ell=0}
     C_\ell (\log (t+i0))^\ell \right) &\hbox{if} \ \alpha\neq-1,-2\cr
               t^{\alpha+2}\left( \sum^{j-1}_{\ell =0}
            C_\ell (\log (t + i0))^{\ell +1} \right)
        &\hbox{if} \ \alpha =- 1\ {\hbox{or}}\ -2}.\eqno(4.4)
$$

(ii) Let $n$ be a non-negative integer and let $I(n, j;t) (j\geq 1)$ denote the
following integral:
$$
\int^\delta_0 {d\rho_1\over \rho_1} \int^{\rho_1}_0 {d\rho_2\over \rho_2}
\cdots
 \int^{\rho_{j-2}}_0
{d\rho_{j-1}\over \rho_{j-1}} \int^{\rho_{j-1}}_0 (t+\rho_j)^n
 \log (t+\rho_j + i0)d\rho_j.
$$
Then its singularity near $t=0$ is of the following form with some constants
$C_\ell(\ell =0, \cdots, j-1)$:
$$
t^{n+1} \left( \sum^{j-1}_{\ell=0} C_\ell (\log (t +i0))^{\ell +1}\right).
\eqno(4.5)
$$

(iii) Let $\widetilde{I}(n,j;t) (n$; a non-negative integer, and $j \geq 1$)
denote the following integral:
$$
\int^\delta_0 {d\rho_1\over \rho_1} \int^{\rho_1}_0 {d\rho_2\over \rho_2}
\cdots \int^{\rho_{j-2}}_0
{d\rho_{j-1}\over \rho_{j-1}} \int^{\rho_{j-1}}_0 \rho_j
(t+\rho_j)^n \log (t+\rho_j+i0)d\rho_j.
$$
Then its singularity near $t=0$ is of the following form with some constants
$C_\ell (\ell =0, \cdots, j-1)$:
$$
t^{n+2} \left( \sum^{j-1}_{\ell =0} C_\ell (\log (t +i0))^{\ell +1}\right).
\eqno(4.6)
$$

\medskip

\noindent{\bf Proof.} Since $\rho_j(t+\rho_j)^\alpha = (t + \rho_j)^{\alpha +
1} -
 t(t+\rho_j)^\alpha$,
 (i) and (iii) follow respectively from Proposition 4.1 and from
 (ii) above.
Hence it remains to prove (ii).
Since
$$
{d^{n+1}\over dt^{n+1}} \left( (t+\rho_j)^n \log (t+ \rho_j+i0)\right) =
{n!\over
t+\rho_j+i0} + P_n,
$$
where $P_n$ is a polynomial of $(t+\rho_j)$, Remark 4.1,
entails that
$$
{d^{n+1}\over dt^{n+1}} I(n,j;t) + n!J(-1, j;t) + h(t) \eqno(4.7)
$$
holds with a holomorphic function $h(t)$.
On the other hand, near $t=0$  a straightforward computation shows
$$
\int^t dt\ t^n (\log (t+i0))^m = {t^{n+1}\over n+1} \sum^m_{r=0} {(-1)^rm!(\log
(t+i0)^{m-r}\over (m-r)!(n+1)^r} \eqno(4.8)
$$
holds for non-negative integers $n$, and $m\neg 1$.
Combining (4.7) and Proposition 4.1,
 we use (4.8) repeatedly to
find (4.5). We also note that Remark 4.2 entails $C_{j-1} =(-1)^j/j!(n+1)$
in this case.

The following proposition is a key result of this section.

\noindent{\bf Proposition 4.3.} Let $I(t)$ denote the following integral
(4.9), where $e_j\ (j=1,2,\cdots ,n)$ is a non-negative integer:
$$
\int^\delta_0 r^{e_1}_1dr_1 \int^1_0 r_2^{e_2} dr_2 \cdots \int^1_0 r_n^{e_n}
dr_n
\log (t+ r_1 \cdots r_n + i0).\eqno(4.9)
$$
Then its singularity near $t=0$ is a sum of finitely many  terms of the form
$$
Ct^N(\log (t+ i0))^m\eqno(4.10)
$$
with a constant $C$ and positive integers $N(\geq min\ \  e_j + 1)$ and $m(\leq
n)$.

{\ul{Proof.}} First of all, let us re-scale the parameter $r_1$ and the
variable
$t$ as follows:
$$
r_1 = \delta r'_1, \ \  t= \delta s
$$
Then $I$ becomes
$$
\delta^{-e_1-1} \int^1_0 (r'_1)^{e_1} dr'_1 \int^1_0 r_2^{e_2} dr_2 \cdots
 \int^1_0 r_n^{e_n} dr_n (\log (s+ r'_1 r_2 \cdots r_n) + \log
\delta). \eqno(4.11)
$$
The contribution from $\log \delta$ in (4.11) is a finite constant.
Thus we may assume from the first that $\delta =1$.
Then the roles of $r_j's$ in (4.9) are uniform, and hence we may re-number the
index $j$ so that
$$
e_1 \geq e_2 \geq \cdots \geq e_n. \eqno(4.12)
$$
Let us introduce new variables $\sigma_j$ by
$$
\sigma_1 = r_1, \ \ \sigma_2=r_1r_2,\cdots,\ \sigma_n = r_1 \cdots r_n
.\eqno(4.13)
$$
The integral $I$ (with $\delta =1$) can be now expressed as
$$
\int^1_0 {\sigma_1^{d_1}\over \sigma_1} d\sigma_1 \int^{\sigma_1}_0
{\sigma_2^{d_2}\over \sigma_2} d\sigma_2 \cdots \int^{\sigma_{n-2}}_0
{\sigma_{n-1}^{d_{n-1}}\over \sigma_{n-1}} d\sigma_{n-1} \int^{\sigma_{n-1}}_0
\sigma_n^{e_n} d\sigma_n \log (t + \sigma_n + i0), \eqno(4.14)
$$
where $d_j = e_j - e_{j+1}$.
The number $d_j$ is nonnegative by (4.12), and this non-negativity
makes our reasoning much simpler: that is why we
re-numbered the index $j$.

The first integration in (4.14),
i.e.
$\int^{\sigma_{n-1}}_0 \sigma^{e_n}_n d\sigma_n \log (t+\sigma_n + i0)$,
can be done in
a straightforward manner: using the identity
$\sigma^e = \sum^e_{j=0} c_j t^j (t+\sigma)^{e-j}$, where $c_j$ is some
constant, we find it is a sum of terms of the form
$$
\eqalignno{
C t^j \big\{ (t &+ \sigma_{n-1})^{e_n-j+1} \log (t + \sigma_{n-1} +i0)\cr
         &- t^{e_n-j+1} \log  (t + i0)\big\}&(4.15)\cr}
$$
and polynomials of the form
$$
C't^j \{ (t+\sigma_{n-1})^{e_n-j+1} - t^{e_n-j+1}\},
$$
where $C$ and $C'$ are some constants.

If $d_{n-1} \geq 1$, the same computation can be done for the second
integration
(4.14).
In this case we do not need to combine the first term and the second
term in (4.15). That is, we perform the integration of these terms separately.
If $d_{n-1}$ is equal to $0$, we first define an integral $J$ by
$$
\int_0^1 {\sigma^{d_1}\over \sigma_1} d\sigma_1 \int^{\sigma_1}_0
{\sigma_2^{d_2}\over \sigma_2} d\sigma_2 \cdots \int^{\sigma_{n-2}}_0
{d\sigma_{n-1}\over \sigma_{n-1}} \{ (t+\sigma_{n-1})^\alpha \log
(t+\sigma_{n-1}+i0) - t^\alpha \log (t + i0)\},
$$
where $\alpha = e_n - j+1$ is a positive integer.
Then we have
$$
\eqalignno{
(t {d\over dt} &- \alpha )J(t)\cr
               &= \int^1_0 {\sigma_1^{d_1}\over \sigma_1} d\sigma_1
               \int^{\sigma_1}_0{\sigma_2^{d_2}\over \sigma_2} d\sigma_2\cdots
               \int^{\sigma_{n-2}}_0
               d\sigma_{n-1} \big\{ - \alpha (t + \sigma_{n-1})
               ^{\alpha -1} \log (t + \sigma_{n-1}+i0)\cr
               &+ ( (t + \sigma_{n-1})^{\alpha -1} - t^{\alpha -1})
               /\sigma_{n-1}\big\}
\cr}
$$
Since the contribution from $((t + \sigma_{n-1})^{\alpha -1} - t^{\alpha -1})
/\sigma_{n-1}$
is finite and analytic in $t$ (actually a polynomial),
the main contribution to $J(t)$ is from $-\alpha (t + \sigma_{n-1})^{\alpha -1}
\log (t + \sigma_{n-1} +i0)$.
But this is the same integral discussed at the first step. Repeating this
procedure we finally find that $I$ has the form $\sum^{e_n}_{j=0} t^jI_j$,
where
$I_j$ satisfies the following equation:
$$
\prod^{n(j)}_{\ell=1} (t {d\over dt} - \alpha_\ell (j) ) I_j (t) =
\sum_k (C_k\log (t + i0) + C'_k) t^k + A, \eqno(4.17)
$$
where $A$ is analytic at $t=0$.
Here $n(j) \leq n-1, \ \ \alpha_\ell (j)$ is an integer $\geq e_n - j+1, k $
ranges over a finite subset of integers $\geq e_n - j+1$, and $C_k$ and
 $ C'_k$ are constants. As a solution of the equation (4.17), $I_j(t)$ [modulo
a function analytic at
$t=0$] is a sum of terms of the form
$$
C t^N (\log (t + i0))^m
$$
with a constant $C$ and integers $N\geq e_n-j+1$ and $m \leq n$.
Thus $I(t)$ consists of terms of the required form (4.10).
Note that min $e_j = e_n$ by the re-numbering of the index $j$.

\medskip

\noindent{\bf Remark 4.3.}
 (i) If log $(t+ r_1 ... r_n + i0)$ in (4.9)
is replaced by $(t + r_1 ... r_n + i0)^\alpha$ ($\alpha$: non-integer), the
resulting integral is a finite sum of terms of the form
$$
Ct^{\alpha+e} (\log (t+i0))^m\eqno(4.10')
$$
with an integer $e\geq \min e_j +1$ and a non-negative integer $m\leq n-1$.
If $\alpha =-1$, then the condition on $e$ is the same as above but the
condition on $m$ is replaced by $m\leq n$.

(ii)  Let $a(r')$ be an analytic function of $r'= (r'_1 ..., r'_n$), in a
closed
``cube'' $C=[\epsilon, 1] \times ... \times [\epsilon, 1] (\epsilon > 0)$.
Then the following integral $F(a)$ has the same singularity as $I(t)$,
or a weaker one :
$$
F(a) = \int^1_\epsilon dr'_1 ... \int^1_\epsilon dr'_{n'} a(r') \int^\epsilon_0
r_1^{e_1} dr_1 ... \int^\epsilon_0 r_n^{e_n} dr_n \log (t+r'_1 ... r'_{n'} ...
r_n + i0).
$$
In fact, for $r'$ in $C$ we find
$$
\eqalignno{
&\int^\epsilon_0 r_1^{e_1} dr_1 ... \int^\epsilon_0 r_n^{e_n} dr_n \log (t +
r'_1 ... r'_m r_1 ... r_n + i0)\cr
= &\int^\epsilon_0 r_1^{e_1} dr_1 ... \int^\epsilon_0 r_n^{e_n} dr_n \big\{
\log
({t\over r'_1 ... r'_m} + r_1 ... r_n + i0)\cr
&+ \log (r'_1 ... r'_m)\big\}.\cr}
$$
Since the contribution from $\log (r'_1 ... r'_m)$ to $F(a)$ is an analytic
function, it suffices to consider the contribution from $\log ({ t\over r'_1
...
r'_{n'}} + r_1 ... r_n + i0)$.
Proposition 4.3 then tells us that it is a sum of terms of the form
$$
\int^1_\epsilon dr'_1 ... \int^1_\epsilon dr'_{n'} a(r') {t^N\over (r'_1 ...
r'_{n'})^N} (\log (t + i0) - \log (r'_i ... r'_{n'}))^m.
$$
Hence the singularity of $F(a)$ near $t=0$ is a sum of terms of the form
(4.10).
Note that the effect of changing the upper end-point of the integral in (4.9)
to
$\epsilon$ is absorbed by the harmless change of scaling in $r$ variables and
$t$ variable (as was employed at the beginning of the proof of Proposition
4.3),
on the condition that $\epsilon$ is a fixed positive constant.

To generalize Proposition 4.3 to the form needed in section 6 we prepare the
following Lemma:

\noindent{\bf Lemma 4.1.}  Let $L_m(t; a) (n =1, 2, ...; {a}$ a strictly
positive constant), denote the following integral:
$$
\int^a_0 {(\log w)^m\over t + w + i0} dw.
$$
Then the singularity structure of $L_m$ near $t=0$ is as follows:
$$
L_m = \sum^{m+1}_{j=1} C_j (\log (t+i0))^j + h(t)\eqno(4.18)
$$
where $C_j (j=1, ..., m+1)$ are some $a$-dependent constants and $h(t)$ is an
a-dependent holomorphic function near $t=0$.

\noindent{\bf Proof.} Since $(\log w)^m \theta (w) \theta (a- w)
(a > 0)$ is well-defined as a product of locally summable functions, the
convolution-type integral $L_m(t; a)$ is well-defined, and it is a boundary
value of a
holomorphic function on $\{\Im t >0\} $ near $t=0$.
To find out its explicit form
(4.18), we first apply an integration by parts:
 $$
\eqalignno{
L_m &= {\mathop{\lim}_{\kappa\downarrow 0}}
       \Bigg\{ - \int^a_\kappa
       {m (\log (w + i0))^{m-1}\over w +i0}
       \log (t + w +i0) dw\cr
    &+ (\log a)^m \log (a+t+i0) - (\log (\kappa + i0))^m
       \log (t+\kappa +i0)
        \Bigg\}\cr
    &={\mathop{\lim}_{\kappa\downarrow 0}} \Bigg\{
      -\int^a_\kappa {m(\log (w +i0))^{m-1}\over w+i0}
       \log \left( {t + w +i0\over t +i0} \right) dw\cr
    &- \left( \int^a_\kappa {m (\log (w +i0))^{m-1}\over w+i0} dw\right)
       \log (t +i0) -
      \left(\log (\kappa + i0))^m  \log (t + \kappa + i0)\right\}\cr
    &+ (\log a)^m \log (a + t + i0)\cr
    &= {\mathop{\lim}_{\kappa\downarrow 0}}
      \Bigg\{ - \int^a_\kappa {m(\log(w + i0))^{m-1}\over w+i0}
      \log \left( {t + w+i0\over t+i0}\right) dw\cr
    &- (\log (\kappa + i0))^m \log { t+\kappa +i0\over t
       + i0}\Bigg\} + (\log a)^m
    \log {t +a+i0\over t+i0}&(4.19)\cr}
$$
Let us note that, if Im $t>0$,
$$
\eqalignno{
&(\log (\kappa + i0))^m \log {t + \kappa + i0\over t+i0}\cr
=&(\log (\kappa + i0))^m \left( { \kappa + i0\over t+i0} - \half \left({ \kappa
+ i0\over t+ i0}\right)^2 + ... \right) \longrightarrow 0\cr}
$$
as $\kappa\downarrow 0$. Hence we obtain
$$
L_m =M_m - (\log a)^m \log (t +i0)+ ( \log a)^m \log (t+a+i0),\eqno(4.20)
$$
where
$$
M_m = {\mathop{\lim}_{\kappa\downarrow 0}} \left( - \int^a_\kappa {m(\log
(w+i0))
^{m-1}\over w+i0}\log \ {t+w+i0\over t+i0} dw\right).\eqno(4.21)
$$
Since $(\log\ (w+i0))^{m-1}$ is locally summable, the reasoning used to verify
the well-definedness of the integral $\displaystyle{  \int^\delta_0 {dr\over r}
\log
{t+r+i0\over t+i0}}$ (cf. the beginning of this appendix) is applicable also to
$M_m$.
To find out the explicit form of $M_m$, let us first note
$$
M_1 = C_2 (\log (t + i0))^2 + C_1 (\log (t +i0)) + h(t)
$$
holds near $t=0$ for some constants $C_1, C_2$ and some holomorphic function
$h(t)$. (Cf. (4.2)). Thus we can verify (4.18) for $n=1$.
For $n > 1$, we use  mathematical induction: Let us suppose (4.18) is
verified for $1 \leq m \leq m_0$.
Since
$$
t {d\over dt} M_{m_0 + 1} = \int^a_0 {(m_0+1)(\log (w + i0))^{m_0}\over t+w
+i0}
dw = (m_0 +1) L_{m_0}, \eqno(4.22)
$$
using the induction hypothesis, we find that $(t {d\over dt})^{m_0 +1} L_{m_0}$
is holomorphic near $t=0$.
This means that $(t{d\over dt})^{m_0+2} M_{m_0+1}$ is holomorphic near $t=0$.
Otherwise stated,
$$
M_{m_0+1} = \sum^{m_0+2}_{j=1} \widetilde{C}_j (\log (t+i0))^j +
\widetilde{h}(t)
$$
holds near $t=0$ for some constants $\widetilde{C}_j (j=1, ..., m_0 + 2)$
and some holomorphic function $\widetilde{h} (t)$.
Therefore (4.20) implies that (4.18) is true for $m =m_0+1$.
Thus the induction proceeds.

\medskip

\noindent{\bf Proposition 4.4.} (i) Let $K_{n,m} (t) (n, m= 1,2,3, ...)$
denote the following integral (with $\delta >0)$ :
$$
\int^\delta_0 (\log r_0)^m dr_0 \int^1_0 ... \int^1_0 \prod^n_{j=1} dr_j
(t+r_0r_1 ... r_n +i0)^{-1}.\eqno(4.23)
$$
Then the singularity structure of $K_{n,m}(t)$ near $t=0$ is as follows:
$$
K_{n,m}(t) =\sum^{n+m+1}_{j=1} C_j (\log (t+i0))^j + h(t),\eqno(4.24)
$$
where $C_j (j=1, ... , n+m+1)$ are some constants and $h(t)$ is some
holomorphic function near $t=0$.

(ii) Let $J_{n,m}(t) (n, m = 1,2,3, ...)$ denote the following integral:
$$
\int^\delta_0 (\log r_0)^m dr_0\int^1_0 ... \int^1_0 \prod^n_{j=1} dr_j \log (t
+r_0 r_1 ... r_n + i0).\eqno(4.25)
$$
Then the singularity structure of $J_{n,m}(t)$ near $t=0$ is as follows:
$$
J_{n,m}(t) = \sum^{n+m+1}_{j=1} C_j t(\log (t+i0))^j + h (t), \eqno(4.26)
$$
where $C_j (j=1, ... , n+m+1)$ are some constants and $h(t)$ is some
holomorphic
function near $t=0$.

\noindent{\bf Proof}. (i) Let $\rho_j (j=0, 1, ..., n)$ denote $\prod^j_{i=0}
r_i$.
Then $K_{n,m}$ assumes the following form:
$$
\int^\delta_0 {(\log \rho_0)^m\over \rho_0}
d\rho_0 \int^{\rho_0}_0
{d\rho_1\over \rho_1} ... \int^{\rho_{n-2}}_0 {d\rho_{n-1}\over \rho_{n-1}}
\int^{\rho_{n-1}}_0 d\rho_n (t+\rho_n+i0)^{-1}.
$$
Hence we find
$$
(- t{d\over dt})^n K_{n,m}(t) = \int^\delta_0 {(\log \rho_0)^m\over t +
\rho_0+i0} d\rho_0.
$$
Therefore Lemma 4.1 shows that $(-t{d\over dt})^{n+m+1}K_{n,m}(t)$ is
holomorphic near $t=0$.
This entails (4.24).

(ii) Since ${d\over dt} J_{n,m}= K_{n,m}$, the result (i) entails
$$
{d\over dt} J_n = \sum^{n+m+1}_{j=1} C_j (\log (t + i0))^j + h(t)\eqno(4.27)
$$
holds for some constants $C_j$ and a holomorphic function $h(t)$.
Hence, by integrating both sides of (4.27), we find (4.26).
Here we have used a formula
$$
\int^t (\log t)^N dt = t\left(\sum^N_{\ell =0} (-1)^\ell {N!\over (N-\ell)!}
(\log t)^{N-\ell}\right).
$$

\medskip

The following generalization of Proposition 4.4 is used in section 6

\noindent{\bf Proposition 4.5.} Let $L_{n,m}(t)(n,m=1,2,3,...)$ denote the
following integral (with $\delta_0 > 0)$, where $e_j(j=0, 1, ..., n)$ is a
non-negative integer:
$$
\int^{\delta_0}_0 (\log r_0)^m r_0^{e_0} dr_0 \int^1_0 ... \int^1_0
\prod^n_{j=1}
r_j^{e_j} dr_j \log (t+r_0r_1r_2 ... r_n + i0).
$$
Then the singular part of $L_{n,m}(t)$ near $t=0$ is a finite sum of terms of
the following form:
$$
C_{N,p}\ t^N(\log (t+i0))^p,\eqno(4.28)
$$
where $C_{N,p}$ is a constant, $N$ is a non-negative integer $(\geq
\displaystyle{\mathop{\min}_{0\leq j\leq n}} e_j +1)$ and $p$ is a positive
integer $(\leq
n+m+1)$.

\noindent{\bf Proof.} Making use of the scaling transformation of $r_0$ and $t$
as in the proof of Proposition 4.3, we may assume without loss of generality
that
$\delta_0=1$.
Furthermore, as the role of variables $r_j(j=1, ... , n)$ is uniform, we may
assume, by re-labelling of the variables $r_j(j=1, ..., n)$, that $e_1\geq e_2
\geq ... \geq e_n$.
If $e_0 \geq e_1$ then the method used in the proof of Proposition 4.3,
supplemented by Lemma 4.1, establishes the required result.
However, this condition cannot be expected to hold in general, and hence we
must generalize. Introducing the new variables $\sigma_j = r_0r_1 .. r_j\ (j=0,
1, ...,
n)$
we find
$$\eqalignno{
L_{n,m}(t) &= \int^1_0 (\log \sigma_0)^m \sigma_0^{d_0-1}d\sigma_0
\int^{\sigma_0}_0 \sigma_1^{d_1-1}d\sigma_1 ...
\int^{\sigma_{n-2}}_0\sigma_{n-1}^{d_{n-1}-1} d\sigma_{n-1}\cr
&\times \int^{\sigma_{n-1}}_0
\sigma_n^{e_n} \log (t + \sigma_n+i0) d\sigma_n,\cr}
$$
where $d_j = e_j - e_{j+1}$.
As noted above, the proof is finished if  $d_0 \geq 0$.
Let us consider the case $d_0 < 0$.
We then use mathematical induction on $m$.
When $m=1$, we use the following:
$$
\eqalignno{
{d\over d\sigma_0}&((\log \sigma_0)\sigma^{d_0}_0 F(\sigma_0, t))\cr
&= d_0 (\log \sigma_0) \sigma_0^{d_0-1} F(\sigma_0, t) + \sigma_0^{d_0-1}
F(\sigma_0, t)\cr
&+ (\log \sigma_0) \sigma_0^{d_0} {\partial F(\sigma_0, t)\over
\partial\sigma_0}.&(4.29)\cr}
$$
Choosing
$$
\int^{\sigma_0}_0 \sigma_1^{d_1-1} d\sigma_1 ... \int^{\sigma_{n-2}}_0
\sigma_{n-1}^{d_{n-1}-1} d\sigma_{n-1} \int^{\sigma_{n-1}}_0 \sigma_n^{e_n}
\log
(t + \sigma_n + i0) d\sigma_n
$$
as $F(\sigma_0,t)$, we obtain
$$
\eqalignno{
d_0 L_{n,1} &= (\log \sigma_0) \sigma_0^{d_0} F(\sigma_0,t)
|_{\sigma_0 =1}-\mathop{\lim}_{\sigma_0\downarrow 0} ((\log \sigma_0)
\sigma_0^{d_0} F(\sigma_0, t))\cr
&- \int^1_0 \sigma_0^{d_0-1} F(\sigma_0,t) d\sigma_0 - \int^1_0 (\log \sigma_0)
\sigma_0^{e_0-e_1}\sigma_0^{e_1-e_2-1}\cr
&\times \int^{\sigma_0}_0 \sigma_2^{d_2-1} d\sigma_2 ... \int^{\sigma_{n-1}}_0
\sigma_n^{e_n} \log (t + \sigma_n + i0) d\sigma_n.&(4.30)\cr}
$$
For  notational convenience, let $A_j (j=1,2,3,4)$ denote the $j$-th term in
RHS of (4.30).
Since $F(1, t)$ is a well-defined integral (cf. Proposition 4.3), $A_1$
vanishes
because of the trivial fact $\log 1=0$.
To confirm that $A_2$ also vanishes, we note that
$$
\abs{ {1\over t+\sigma_n}} \leq C_\epsilon
$$
holds if $\Im t \geq \epsilon >0$ and $\sigma_n$ is real. Then we find, for
$\sigma_0 \geq 0,$
$$
\eqalignno{
|F(\sigma_0, t) | &\leq C_\epsilon \int^{\sigma_0}_0 \sigma_1^{d_1-1} d\sigma_1
... \int^{\sigma_{n-1}}_0 \sigma_n^{e_n} d\sigma_n\cr
&= {C_\epsilon \sigma_0^{e_1+1}\over \prod^n_{j=1} (e_j +1)}.
\cr}
$$
Therefore
$$
\abs{A_2} \leq C_\epsilon \mathop{\lim}_{\sigma_0\downarrow 0} {(\log \sigma_0)
\sigma_0^{e_0+1}\over \prod^n_{j=1} (e_j +1)}=0.
$$
Since $\epsilon$ is an arbitrary positive number, this means that $A_2$
vanishes.

The term $A_3$ has the same structure as the integral discussed in Proposition
4.3, and hence its singular part is a sum of terms of the form (4.28).
Note that we can re-label all variables including $r_0$ if we go back to
$r$-variables from $\sigma$-variables in the integral $A_3$; the factor $\log
\sigma_0$ has disappeared in $A_3$.

Finally let us study $A_4$. As it has the form $L_{n-1,1}$, we can apply the
above procedure to it.
Repeating this procedure , we eventually end up with one of the following two
integrals (i) or (ii), together with terms of the form $(4.28)$:

(i) $L_{n',1} (n' < n) $ with $d_0 \geq 0$

(ii) $\int^1_0 (\log \sigma_0) \sigma_0^{e_0+1} \log (t+\sigma_0+i0)
d\sigma_0$.

\noindent By using Lemma 4.1 together with the method of the proof of
Proposition 4.3, we can verify that  the singular part of either of them is a
sum of
terms of the form (4.28).

Thus the proof is finished if $m=1$.
Let us consider next the case $m \geq 2$.
We then use
$$
\eqalignno{
{d\over d\sigma_0} &(( \log \sigma_0)^m \sigma_0^{d_0} F(\sigma_0, t))\cr
&= d_0 (\log \sigma_0)^m \sigma_0^{d_0-1} F(\sigma_0,t)+ m(\log \sigma_0)
^{m-1}\sigma_0^{d_0-1}F(\sigma_0, t)\cr
&+ (\log \sigma_0)^m \sigma_0^{d_0} {\partial F(\sigma_0, t)\over \partial
\sigma_0}.&(4.31)\cr}
$$
As  before, we concentrate our attention on the case $d_0 <0$.
Choosing as $F(\sigma_0, t)$ the same integral as was used when $m=1$, we find
that the same reasoning as before applies to the contribution from the
LHS of $(4.31)$ and the third term on the RHS of $(4.31)$.
When integrated over [0, 1] (with respect to $\sigma_0)$,
the second term on the RHS of $(4.31)$ turns out to be $mL_{n,m-1}$.
Thus the induction proceeds, completing the proof.
\newpage
\noindent{\bf 5. Weakness of the singularity in the
general  non-separable meromorphic case.}
\medskip

To confirm the weakness of the singularity in the non-separable meromorphic
case we first need to verify
$$
{\partial\over \partial \rho_i } \varphi (q -\Delta)|_{\varphi =0} \neq 0,
\eqno(5.1)
$$
where $\rho_i = r_1 \cdots r_i$, with $i$ being the index labelling the first
{\it bridge} line; i.e., $i$ is the smallest $j$ such that the photon line $j$
has a meromorphic coupling on both ends, and completes to a closed loop ---
constructed according to the rules specified below Eq.(2) in Ref. 2  --- that
flows
along at least one $*$-segment. The $k_i$-dependent vector $\Delta$ is chosen
so that at {$\varphi =0$} the pole factor associated with each $*$-segment can
be evaluated at the critical point $p_s(q-\Delta) \ (s = 1,2,3)$, defined below
Fig. 1 in Ref. 2, with $q=(q_1,q_2,q_3)$ the set of external variables defined
there.

The vector $\Delta$ is constructed in the following way.
Introduce for each bridge line $i$ an open flow line
$L(k_i)$ that passes along
this photon line $i$, but along no other photon line, and along no
$*$-segment. Instead, the flow line $L(k_i)$ {\it enters} the diagram at one
of the three vertices
$v_i$ and {\it leaves} at another. Specifically, let $e$ be an end-point of the
photon line $i$, and let $s$ be the side of the triangle
on which $e$ lies. This point $e$ separates $s$ into two connected components,
$s^0$
and $s^*$, where $s^*$ is the part of $s$ that contains the $*$-segment.
Run $L(k_i)$ along the component $s^0$. At the end-point of $s^0$ that
coincides with a vertex $v_i$ of the triangle
diagram, run  $L(k_i)$ out along the external line $q_j\ (j=1, 2 $ or 3).
Do the same for the other end-point of the line $i$.
Include on $L(k_i)$ also the segment  $i$ itself.
This produces a continuous flow line. Orient it so that it agrees
with the orientation of the line $i$.
This oriented line is the flow line $L(k_i)$.
Then for each external line $q_j$ along which $L(k_i)$ runs add to the vector
$q_j$ either $+k_i$ or $- k_i$ according to
whether the orientation of $L(k_i)$ is the same as, or opposite to, the
orientation of the external line $q_j$ along which $L(k_i)$ runs.
Sum up the contributions from all of the bridge lines. This shift in
$q=(q_1,q_2,q_3)$ is the vector $\Delta$.

\newpage

The function of interest has the form
$$\eqalignno{
F(q) &=\prod^n_{j=1}\int_{\Omega_j \widetilde{\Omega}_j=1} d\Omega_j
\prod^n_{j=i+1} \int^1_0 r_j^{e_j}dr_j
\prod^i_{j=1} \int^1_0 r_j^{e_j}dr_j\cr
&\times A(q, \Omega, r) \log \varphi(q-\Delta),&(5.2)\cr}
$$
where
$$
D \log \varphi(q-\Delta)+E = \int_{p\approx \widehat{p}} d^4p \prod^3_{s=1}
{1\over p^2_s - m^2+i0},\eqno(5.3)
$$ and $A$, $D$, and $E$ are holomorphic.

Here
$$\eqalignno{
p_1 &= p+ q_1 + \sum^n_{m=i} \epsilon_{1m} k_m,\cr
p_2 &= p - q_3 + \sum^n_{m=i} \epsilon_{2m} k_m,&(5.4)\cr}
$$
and
$$
p_3 = p + \sum^n_{m=i} \epsilon_{3m} k_m,
$$
with each $\epsilon_{sm}$ either zero or one.

We are interested in the singularity of this function at the point
$\widehat{q}$
on $\varphi (q) =0$.
This singularity comes from the $p$-space point $\widehat{p} = p(\widehat{q})$,
 and we can consider the $p$-space domain of
integration to be some small neighborhood of $\widehat{p}$.
Similarily, the domain in $(r,\Omega)$ is confined to a
region $R$ in which the following conditions hold:
$$\eqalignno{
(\widehat{p} + \widehat{q}_1) \cdot &\sum^n_{m=1} \epsilon_{1m} k_m/\rho_i
\approx
i\epsilon_1\cr
(\widehat{p} - \widehat{q}_3) \cdot &\sum^n_{m=1} \epsilon_{2m}
k_m/\rho_i\approx
i\epsilon_2\cr
\widehat{p} \cdot &\sum^n_{m=1} \epsilon_{3m} k_m/\rho_i \approx
i\epsilon_3.&(5.5)\cr}
$$
That is, the real parts of the three denominators in $(5.3)$ are close to zero,
and the imaginary
parts are positive: $\epsilon_s \geq 0 (s = 1,2,3); \sum \epsilon_s >0$.
It was shown in Ref. 2 that  the contours can be distorted in a way such that
(5.5)
holds in a neighborhood of the points contributing to
the singularity at $\widehat{q}$.

Note that all of the $k_m$ that contribute to $(5.5)$ belong to bridge lines,
and hence have a factor $\rho_i$. Thus none of the $r_j(j\leq i)$ enter into
$(5.5)$. Hence the region $R$ is independent of the variables $r_j(j \leq i)$.

The quantity $\Delta = (\Delta_1, \Delta_2, \Delta_3)$ is added to $q = (q_1,
q_2, q_3)$,
and it satisfies, in analogy to $\sum q_i =0$, the condition $\sum \Delta_i
=0$.
This trivector $\Delta$ is a sum of terms, one for each bridge line.
For each bridge line $j$ the corresponding term in $\Delta$ is proportional to
$k_j$.
If line $j$ bridges over (only) the star line on side $s=1$ then the
contribution to $\Delta$
is $(- k_j, k_j, 0)$.
If line $j$ bridges over (only) the star line on side $s=2$ then the
contribution to $\Delta$ is $(0, -k_j, k_j)$.
If the line $j$ bridge (only) over the star line on side $s=3$ then the
contribution to $\Delta$ is
$(k_j, 0, - k_j)$.

The gradient of $\varphi (q)$ is also a trivector.
The condition $\sum q_i =0$ in $q$ space means that the gradient (which is in
the dual space) is defined modulo translations: $\Delta_i \rightarrow \Delta_i
+ X$, all $i$.
Thus one can take $\nabla\varphi$ to have a null second component.
Then at the point $\widehat{q}$ of interest the gradient has the form$^{3}$
$$
\nabla \varphi = (\alpha_1\widehat{p}_1, 0, - \alpha_2\widehat{p}_2),\eqno(5.6)
$$
provided the sign and normalization of $\varphi$ are appropriately defined.
Hence the quantity on the left-hand side of (5.1) is, at $q-\Delta =
\widehat{q}$,
$$\eqalignno{
{\partial\varphi (q-\Delta)\over \partial \rho_i} &= - \nabla \varphi \cdot
{\partial\Delta\over \partial\rho_i}\cr
&= \sum^3_{s=1} \sum^n_{m=i} \alpha_s\widehat{p}_s \epsilon_{sm} k_m/\rho_i
&(5.7)
\cr}$$
which, according to (5.5), is nonzero, as claimed in (5.1).
Use has been made here of the Landau equation $\sum \alpha_s\widehat{p}_s=0$.

Using (5.1) we now employ the result in section 3 to normalize the defining
function
 $\varphi$ of the Landau surface so that we may
apply Proposition 4.3 of
Section 4 to the integral $F$ in question.
It follows from Lemma 3.1 in Appendix 3 that the following normalization holds
on a neighborhood
of the point in question:
$$
\varphi (q-\Delta) = B(q,\rho_i, k'/\rho_i) \ (\rho_i - \varphi (q)/C(q,
k'/\rho_i )), \eqno(5.8)
$$
where $B$ and $C$ are different from 0 at any point in question, and $k'$
denotes the totality of the bridge lines $k_j$.
Note that each bridge $k_j$ contains a factor $\rho_i$ and that $k'/\rho_i$ is
independent of $\rho_i$.
Let us now apply Proposition 4.3 in section 4 to the following integral I:
$$
I= \int^\delta_0 r_1^{e_1} dr_1 \int^1_0 r_2^{e_2} dr_2 \cdots \int^1_0
 r_i^{e_1} dr_i \log (\rho_i - \varphi / C). \eqno(5.9)
$$
Then we find [modulo a function analytic at $\varphi=0$]
$$
\eqalignno{
I = E(q, k'/ \rho_i) \ (\varphi (q)/C(q,l'/\rho_i))^N
\ &\bigg( \sum^i_{j=0} a_j
   (q, k'/\rho_i)\cr
&\times \left( \log (\varphi (q)/C(q, k'/\rho_i))\right)^j \bigg) &(5.10)\cr}
$$
with $N\geq 1$, and $E$ and $a_j$ being holomorphic in their arguments,
and,  in particular, in the
$r_j$'s $(j > i)$.

The function $A$ in $(5.2)$ is holomorphic. This factor has no important effect
on the result: it can be incorporated by using Remark 4.3(ii) of section 4.
\newpage
\noindent{\bf 6. Computation for the Nonmeromorphic Case}

\medskip

The computation in the nonmeromorphic case is similar to the computation for
the
meromorphic case described in section 5 with the help of section 3.
Let the special index $i$ be now the smallest integer such that photon line $i$
is either a bridge line or a photon line with a nonmeromorphic coupling on at
least one end.

If line $i$ is a bridge line (and hence, by definition, has a
meromorphic coupling on each end, and bridges across a $\ast$-segment) then the
argument used for the meromorphic case continues to work.
This is because the condition (5.1) of section 5 continues to hold, and each
variable $\lambda_j$ associated with a nonmeromorphic coupling acts just like a
variable $r_j(j > i)$ of sections 3  and 5.

If, on the other hand, the index $i$ labels a line with a nonmeromorphic
coupling on at least one end then (5.1) may fail, because in this case the
variable $k_i$ may enter into $\Delta$ only in the form $\lambda_i k_i$ (or
$\lambda_i' k_i)$.
For example, if the photon line $i$ runs between two different sides, $s$ and
$s'$, and has a nonmeromorphic coupling on both ends then, according to
(10.8b) of ref. 1, the vector $k_i$ enters into $\Delta$ only in the
combinations $\lambda_ik_i$
or $\lambda_i' k_i$, where $\lambda_i$ and $\lambda_i'$ are the variables
associated with the two different nonmeromorphic couplings of line $i$.
Hence the derivative on $\Delta$ occurring in (5.7) will introduce a factor
$\lambda_i$ or $\lambda'_i$ into each $k_i$-dependent contribution to (5.7).
Since $\lambda_i$ and $\lambda'_i$ vanish in the domain of integration, and all
other contributions have factors $r_j(j >i)$, which can vanish, the property
(5.1) can fail.

Similarly, if only one end of line $i$ is coupled nonmeromophically, say into
the side $s$, but the closed loop $i$ does not pass through the star line for
either of the other two sides $s' \neq s$, then again (5.1) can fail, for
essentially the same reason.

These failures of (5.1) cannot be avoided by simply using $\rho'_i =
\lambda_i\rho_i$ (or $\lambda'_i \rho_i)$ in place of $\rho_i$, because the
condition in (3.1) on $k'/\rho_i$ fails if $\rho_i$ is replaced by $\rho'_i$.

In this section the ``self-energy'' photons that couple nonmeromorphically on
both ends onto the same side $s$ will be ignored: they are treated in
section 7.

To deal with the new cases we introduce the set of variables $x_j(j\epsilon J)$
to represent both the $r_j(j >i)$, and also the occurring variables $\lambda_j
(j\geq i)$ and $\lambda'_j(j\geq i)$.
This set $x_j(j\epsilon J)$ replaces the set $r_j(j > i)$ that occurs in the
arguments of section 3 and 5.

Using the evaluation (5.6) for the constant gradient vector $\nabla\varphi$
we define a new variable
$$
\eqalignno{
\rho &= - \nabla \varphi \cdot \Delta\cr
     &= \sum^3_{s=1}  \sum^n_{m=i} \alpha_s\widehat{p}_s \epsilon'_{sm} k_m,
     &(6.6)\cr}
$$
where the reasoning leading to (5.7) has been used.
However, $\epsilon'_{sm}$ can be 0, 1, $\lambda_m$, or $\lambda'_m$, with the
latter two possibilities coming from the possible nonmeromorphic couplings.

In the case under consideration the photon line $i$ has a nonmeromorphic
coupling on one or both ends.
If this line $i$ has a nonmeromorphic coupling on only one end, and the
$\epsilon'_{si}$ associated with the other end is 1, then (5.1) again holds,
and
the method used in the meromorphic case again works.
In the remaining, cases (namely those for which $\epsilon'_{si}\neq 1$ for all
$s$) the function $r_0 = \rho/\rho_i$ has a term $\lambda_i p_s\Omega_i$ (or
$\lambda'_i p_s \Omega_i)$ and no other dependence on $\lambda_i$ (or
$\lambda'_i$).
Hence the variable $r_0$ may be introduced as a new variable, replacing
$\lambda_i$ (or $\lambda'_i$), provided the associated coefficient $p_s\cdot
\Omega_i$ is nonzero.

The arguments of Ref. 2, slightly extended to include the $\lambda_j$, show
that $\widehat{p}_s\cdot \Omega_i$ can be taken to be nonzero near points in
the integration domain that lead to a singularity of the integral at
$\widehat{q}$.
Hence the transformation to the new set of variables (with $r_0$ replacing
$\lambda_i$ or $\lambda'_i$) is a holomorphic transformation: all analyticity
properties are maintained.

The derivative at $(q - \Delta)=\widehat{q}$ of $\varphi(q-\Delta)$ with
respect
to $\rho$ is
$$
\eqalignno{
{\partial \varphi(q-\Delta)\over \partial\rho} &= \nabla\varphi\cdot
{\partial (-\Delta)\over \partial\rho}\cr
&= {\partial (-\nabla \varphi\cdot \Delta)\over \partial\rho}\cr
&= {\partial\rho\over \partial\rho} =1.&(6.7)\cr}
$$
Thus (5.1) is now satisfied (with $\rho = r_0 r_1 ... r_i$ in place of
$\rho_i=r_1 r_2 ... r_i$), and we can use the method of sections 3 and 5.

The function $F(q)$ of (5.2) now takes the form
$$
F(q) =\prod^n_{j=1} \int_{\Omega_j\widetilde{\Omega}_j=1} d\Omega_j
 \prod^i_{j=1} \int^1_0 r_j^{e_j} dr_j \int dr_0 G(q,r, r_0),\eqno(6.8)
$$
where
$$
\eqalignno{
G(q,r,r_0) &= \prod_{j\epsilon J} \int^1_0 x_j^{e_j} dx_j
\delta(r_0+\rho^{-1}_j\nabla\varphi\cdot\Delta)\cr
&\times A(q,\Omega,r, x) \log \varphi(q-\Delta),&(6.9)\cr}
$$
and $\log \varphi(q-\Delta)$ is defined in (5.3) and (5.4), but with the
$\epsilon'_{sm}$ in place of the $\epsilon_{sm}$.
Notice  that the $\int dr_0$ can be cancelled by the $\delta$ function to give
the generalization of (5.2) engendered by the action of the nonmeromorphic-part
operators of (10.8b) in Ref. 1.

The expression for $G$ given in (6.9) is well defined only for real $\nabla$
(i.e., only for real $k_j(j\geq i))$.
A more general definition is this: (1), leave the $\int dr_0$ and $\delta$
function out of (6.8) and (6.9); (2), change the variable $\lambda_i$ (or
$\lambda'_i$) to
$r_0; (3)$, replace the $\int d\lambda_i$ (or $\int d\lambda'_i)$ by
$\int dr_0$; (4), identify $G$ as the integrand of this integral over $dr_0$.

Near the point $\widehat{q}$ one can write
$$
\eqalignno{
\varphi (q-\Delta) &\cong \varphi (q) - \nabla \varphi \cdot\Delta\cr
&= \varphi(q) + \rho.&(6.10)\cr}
$$
Insertion of (6.10) into (6.8) and (6.9) gives
$$\eqalignno{
F(q) &=\prod^n_{j=1} \int_{\Omega_j\widetilde{\Omega}_j=1} d\Omega_j
 \prod^i_{j=1} \int^1_0 r_j^{e_j} dr_j\int dr_0 \log (\varphi (q)+\rho)
\cr
&\times f(r_0,r,\Omega,q), &(6.11)\cr}
$$
where
$$
f(r_0, r,\Omega, q) = \prod_{j\epsilon J} \int^1_0 x_j^{e_j}
dx_j \delta (r_0 + \nabla\varphi \cdot \widetilde{\Delta})
A(q, \Omega, r, x), \eqno(6.12)
$$
and $\widetilde{\Delta}\equiv \Delta/\rho_i$.

Equation (6.11) exhibits the smearing of the log $\varphi (q)$ singularity.
If $f(r_0,r,\Omega, q)$ were to have a $\delta$ function singularity at
$r_0=0$  then the
expression (6.12) would yield a singularity of the form log $\varphi(q)$.
But if $f$ has only a milder singularity at $r_0=0$ then $F(q)$ will have a
weaker singularity at $\varphi (q) =0$.

\medskip

Let us examine, then, the form of $f(r_0, r, \Omega, q$).
Let the particular $x_j$ that is $\lambda_i$ be called simply $\lambda$.
Then $\nabla\varphi\cdot\widetilde{\Delta}$ will be $(a \lambda + P)$,
 where $P$ is a
sum of terms each of which is a coefficient of the
form $p_s(q)\Omega_j$ times a product $r_{i+1}r_{i+2} ... r_{j}$, or
$r_{i+1}r_{i+2} ... r_{j} \lambda_{j}$, or $r_{i+1}r_{i+2} ...
r_{j}\lambda'_{j}$.
Eventually the coefficients $p_s(q)\Omega_j$ will be shifted to nonzero complex
numbers.
But we shall evaluate the integrals first at points where each $p_s(q)
\Omega_j=1$, $a=A=1$, and each $e_j=0$.

Consider first, then, for $0<r_0<1$, the simple example
$$
f(r_0) = \int^1_0 d\lambda\int^1_0 dx_1 \int^1_0 dx_2\ \delta(r_0 - \lambda -
x_1
x_2).\eqno(6.13)
$$
Using the $\delta$ function to eliminate the $\int d\lambda$ we obtain
(with $\Theta$ the Heaviside function)
$$\eqalignno{
f(r_0) &= \int^1_0 dx_1 \int^1_0 dx_2\ \Theta (r_0 - x_1x_2)\cr
&= \int^1_0 dx_1 \int^{r_0/x_1}_0 dx_2\ \Theta (1 - {r_0\over x_1})\cr
&+ \int^1_0 dx_1 \int^1_0 dx_2\ \Theta ({r_0\over x_1} -1)\cr
&= r_0 \int^1_{r_0} {dx_1\over x_1} + \int^{r_0}_0 dx_1\cr
&= r_0 (-\log r_0 + 1).&(6.14)\cr}
$$
Thus in this case the singularity of the function $f(r_0)$ is much weaker
than $\delta(r_0)$: $f(r_0)$ is bounded and tends to zero as $r_0$ tends to
zero.

The general form of $f(r_0)$ is
$$
f(r_0) = \prod_{j\epsilon J} \int^1_0 dx_j\ \Theta(r_0-P),
\eqno(6.15)
$$
where $P$ is as defined above.
One sees immediately that $f(r_0)$ is bounded, and tends to zero with $r_0$.

To begin the study of the general form of $f(r_0)$ let us consider a case
slightly more complicated than (6.14):
$$\eqalignno{
f(r_0) &= \int^1_0 dx_1 \int^1_0 dx_2 \int^1_0 dx_3\ \Theta(r_0-x_1x_2x_3)\cr
&= \int^1_0 dx_1 \int^1_0 dh\ \Theta(r_0 - x_1 h)\cr
&\times \int^1_0 dx_2 \int^1_0 dx_3\ \delta(h-x_2x_3)\cr
&= \int^1_0 dx_1 \int^1_0 dh\ \Theta (r_0 - x_1 h) H(h),&(6.16)\cr}
$$
where (for $0<h<1$)
$$\eqalignno{
H(h) &= \int^1_0 dx_2\int^1_0 dx_3\ \delta(h-x_2x_3)\cr
&=\int^1_0 {dx_2\over x_2}\ \Theta (1-{h\over x_2})\cr
&= \int^1_h {dx_2\over x_2} = (- \log h).&(6.17)\cr}
$$

Notice that the last line of (6.16) has the same form as the first line of
(6.14), but with a different function $H$. Substituting the function $H(h)$
from
(6.17) into (6.16) one obtains, for $0<r_0<1$,
$$
\eqalignno{
f(r_0) &= \int^1_0 dx_1 \int^1_0 dh\ \Theta ({r_0\over x_1} - h) (-\log h)\cr
&= \int^1_{r_0} dx_1 \int^{r_0/x_1}_0 dh (-\log h)\cr
&+\int^{r_0}_0 dx_1 \int^1_0 dh (-\log h)\cr
&= \sum_{n,m} C_{nm}(r_0)^n (\log r_0)^m,&(6.18)\cr}
$$
where only a finite number of the constant coefficients $C_{nm}$ are nonzero.

A function of one variable $x$ having, in $0<x<1$, the form
$\sum C_{nm} x^n (\log x)^m$, and bounded in $0\leq x \leq 1$,
with some finite number of nonzero coefficients $C_{nm}$, will be said to have
form $F$. Thus the functions $f(r_0)$ specified in $(6.14)$ and $(6.16)$ both
have form $F$.

In fact, the general function $f(r_0)$ of the form specified in $(6.15)$ has
form $F$. To see this note first that if we replace the factor $H(h)=(-\log h)$
in (6.16)
by any function $H(h)$ of form $F$ then $f(r_0)$ has form $F$:
$$\eqalignno{
f(r_0) &= \int^1_{r_0} dx_1 \int^{r_0/x_1}_0 dh\ H(h)\cr
&+ \int^{r_0}_0 dx_1 \int^1_0 dh\ H(h)\cr
&= r_0\int^1_{r_0} {dx'\over(x')^2} \int^{x'}_0 dh\ H(h)\cr
&+ r_0 \int^1_0 dh\ H(h)&(6.19)\cr}
$$
Then $(4.8)$ and Lemma $4.1$ give the result that if $H(h)$ has form $F$ then
$f(r_0)$ has form $F$.
(Note that every term in $\int^{x'}_0 dh H(h)$ has a factor $x'$, and hence the
denominator $(x')^2$ is reduced to $x'$.)
So our problem is to show that $f(r_0)$ can be reduced to the form (6.16) with
$H(h)$ having the form $F$.

To show this let $I(g)$ be some function of form $F$ and consider the integral
operator $H_h$ defined by
$$
H_h[I(g)]= \int^1_0 dx \int^1_0 dg\ \delta (h-xg) I(g).\eqno(6.20)
$$
Then, for $0<h<1$,
$$
\eqalignno{ H_hI &= \int^1_h {dx\over x}\ I({h\over x})\cr
&= \int^1_h {dx'\over x'}\ I(x'),&(6.21)\cr}
$$
where $x'=h/x$. Then $(4.8)$ and Lemma $4.1$ entail that if $I$ has form $F$,
so does $H_hI$.

Repeated application of this result shows that if $P = x_1x_2 ... x_p$ then
$f(r_0)$ has form $F$.
One first combines $x_{p-1} x_p$ into $h_p$, then combines $x_{p-2} h_p$ into
$h_{p-1}$, etc..
At each stage the functions $I$ and $H$ have form $F$, and hence one finally
gets (6.16) with $H(h)$ having form $F$, as required.

The general form of $P$ is not just a single product $r_{i+1} ... r_j$:
it is a sum of such terms with different values of $j$, some of which can be
multiplied by $\lambda_j$ or $\lambda'_j$.
However, these other terms can be brought into the required form by a
generalization of the operator technique used above.

Let us again consider first a simple case:
$$
f(r_0) = \int^1_0 dx \int^1_0 dg \int^1_0 dt \ \Theta (r_0 - xt - x g)
I(g), \eqno(6.22)
$$
where $I(g)$ has form $F$, and $t$ could be a $\lambda_j$.
Then
$$
f(r_0) = \int^1_0 dx \int^2_0 dh\ \Theta(r_0 - x h) H(h),\eqno(6.23)
$$
where
$$
H(h) = \int^1_0 dg \int^1_0 dt\ \delta(h-t-g) I(g).\eqno(6.24)
$$
Thus
$$
\eqalignno{
f(r_0) &= \int^1_0 dx \int^1_0 dh\ \Theta (r_0 - x h) H(h)\cr
&+ \int^1_0 dx \int^2_1 dh\ \Theta (r_0 - x h) H(h).&(6.25)\cr}
$$
For $0<h<1$ the function $H(h)$ is
$$
\eqalignno{H(h) &= \int^1_0 dg \int^1_0 dt\ \delta(h-t-g)I(g)\cr
&= \int^1_0 dg\ I(g) \Theta(h-g)\Theta(1-(h-g))\cr
&= \int^h_0 dg\ I(g),&(6.26)\cr}
$$
which has form $F$.
Thus the first term in (6.25) gives a contribution $f_1(r_0)$ to $f(r_0)$ that
has form $F$.
The second term is, for $0<r_0<1$,
$$\eqalignno{
f_2(r_0) &= \int^1_0 dx \int^2_1 dh\ \Theta(r_0 - xh) H(h)\cr
&= \int^1_0 dx \int^2_1 dh\ \Theta ({r_0\over x} -h)
 \int^1_{h-1} dg\ I(g)\cr
&=\int^1_0 dx \int^{r_0/x}_1 dh\ \Theta ({r_0\over x}-1)
\ \Theta (2-{r_0\over x}) \int^1_{h-1} dg\ I(g)\cr
&+\int^1_0 dx \int^2_1 dh\ \Theta ({r_0\over x} -2) \int^1_{h-1} dg\ I(g)\cr
&= \int^{r_0}_{r_0/2} dx \int^{r_0/x}_1 dh \int^1_{h-1} dg\ I(g)\cr
&+ \int^{r_0/2}_0 dx \int^2_1 dh\ \int^1_{h-1} dg\ I(g)\cr
&= r_0\int^2_1 {dx'\over(x')^2} \int^{x'}_1 dh \int^1_{h-1} dg\ I(g)
+ (r_0/2) \times {\hbox{const.}} \cr
&= r_0 \times {\hbox{const.}},&(6.27)\cr}
$$
which is also of form $F$.

The two important points are: (1), that the integral operator that reduces a
sum
$t+g$ to a single $h$, just like the operator that reduces a product $tg$ to
$h$, preserves form $F$; and (2), the extra part $h>1$ does not
disrupt the argument: it adds only a term $r_0 \times$ const.

By taking combinations of these two kinds of operators, and a third kind with
$t$ fixed at unity, rather than being integrated over, one can reduce any one
of the possible functions $(r_0 -P)$ to $(r_0 - x_1 h)$ combined with an $H(h)$
of form $F$. Thus all functions $f(r_0)$ of the kind $(6.15)$ will be of the
form $F$ , provided we make the simple assignments $1=a=A=p_s\Omega_j=e_j+1$.
The remaining task is to show that essentially the same result follows even
when we do not make these simple assignments. One other problem also needs
to be addressed: we have computed the integrals on the variables $r_j$ under
the assumption that the variables $\Omega_j$ are held fixed, whereas the
distortions in the variables $\Omega_j$ can depend on the $r_j$.

By following through the arguments just given, but with the $e_j$ now
allowed to be nonnegative integers, one finds that the conclusions are not
disrupted: the positive power $n$ of $r_0$ in $f(r_0)$ can be increased, and
the
positive power $m$ of $\log r_0$ can be decreased, but changes in the opposite
direction do not occur. Hence the singularities are at most weakened.

To deal simultaneously with the problems of the dependence of
$A(q,\Omega,r,x)$ upon $(r,x)$, and the dependence of the distortion in
$\Omega$
upon $(r,x)$, we introduce a sufficiently small number $\epsilon = 1/N > 0$,
and divide the domain of integration $0<x<1$ in each of the variables $x_j$
and $r_j$ into a sum of $N$ intervals of length $\epsilon$, such that (1),
the distortion of the set of $\Omega$ variables can be held fixed over each
separate product interval, and (2), for any {\it subset} $\sigma$ of the set of
variables $r_j$ and $x_j$, and for the corresponding space $S$ formed by the
product over $\sigma$ of the corresponding set of {\it leading} intervals
$0<(r_j,x_j)<\epsilon$, the dependence of $A$ on these variables can
be represented by a power series that converges within $S$ for each point in
the space formed by the product over the complementary set of variables of the
nonleading intervals $\epsilon <x<1$. (See Remark 4.3(ii).) The variables can
then
be re-scaled so that the original integration domains run from $0$ to $N$, and
the leading intervals (formerly from $0$ to $\epsilon$) now run from from $0$
to $1$. The earlier arguments can then be applied to the re-scaled problem,
with the concept `form F' replaced by `form F$'$: a function
of one variable $x$ is said to have form F$'$ if and only if it is bounded in
the interval $0\leq x\leq 1$, and in $0<x<1$ can be written in the form
$$
\sum_m A_m(x) (\log x)^m,
$$
where the sum is over a finite set of integers $m$, and each $A_m(x)$ is
analytic
on $0<x<1$. The contributions from the integrations over the nonleading domains
$1<x<N$ do not disrupt the arguments, and formula $(4.8)$ shows that extra
factors $n+1$ are introduced into the denominators at each integration, so that
convergence at the level of the integral is, if anything, improved over the
original convergence at the level of the integrand. This takes care of these
two
problems.

The final step is to remove the assumption that the coefficients of
the various terms of $P$ are unity: these coefficients are actually the
quantities  $p_s\Omega_j$.

There is no problem in allowing these coefficients to be strictly-positive
$\Omega$-dependent functions: the constants $C_{nm}$, or the functions
$A_m$, then simply become analytic functions of the variables $p_s\Omega_j$
over these strictly-positive domains. In fact, these coefficients can be
continued into the complex domain without affecting the character of the
singularity at $r_0=0$ provided we keep each coefficient away from the cut
along the negative real axis in that variable, and keep the point $C$ in the
space of the collection of these coefficients away from all points where
$a\lambda+P(C,x_j,r_j)=0$ for some point in the product of the open domains of
integration $0<\lambda<1$, and (for all $j$) $0<x_j<1$ and $0<r_j<1$.
Here $a=p_s\Omega_i$.

The points in the domain of integration over the variables $r_j,x_j,\Omega_j$
that contribute to the singularity at $\varphi=0$ are points where each of the
three star-line factors is evaluated at, or very close to, the associated pole.
The arguments in ref. 2 show that in this region the first of the variables
$p_s\Omega_j$, namely $p_s\Omega_i=a$ can be shifted into
upper-half plane Im$a>0$, and the collection of contours $C$ can be distorted
so that $a\lambda+P$ is shifted into the upper-half-plane provided
$0<\lambda<1$ and, for all $j$, $0<r_j<1$ and $0<x_j<1$. This is exactly the
condition that is needed to justify the extension of the results obtained
above for positive real coefficients to the complex points of interest.

The dependence of the distortion of the contour on $\lambda$ needs to be
described. When one introduces the nonmeromorphic couplings, and hence the
$\int d\lambda$, into the formula, the Landau matrix acquires a new column,
the $d\lambda$ column. However, the parameter
$\lambda$ enters in an almost trivial way: the pole residues associated
with the side $s$ of the triangle into which the vertex associated
with $\lambda$ is coupled are changed from
$p_s(\Omega_j+ ...)$ into $(p_s+\lambda_i k_i)(\Omega_j+...)$, and the pole
denominator $(p_s)^2 - m^2$ is changed to $(p_s+ \lambda_i k_i)^2 - m^2$.
The new set of Landau equations can be satisfied at each of the two end points
$\lambda_i=1$ and $\lambda_i=0$. These two solutions correspond to diagrams in
which the vertex associated with $\lambda_i$ is placed at one end or the other
of the side $s$ of the triangle. Both solutions to the triangle-diagram
equations
exist, and, because of the null contributions in all $d\Omega_j$ columns, the
two solutions yield two different ways of distorting the $\Omega_j$ contours,
$\Delta_1$ and $\Delta_0$, the first corresponding to $\lambda_i=1$, the second
corresponding to $\lambda_i=0$. An allowed distortion that gives
these two cases and smoothly interpolates to the intermediate
values of  $\lambda_i$ is $\lambda_i \Delta_1 + (1-\lambda_i)\Delta_0$. It
keeps the imaginary part of the pole denominator strictly positive (near the
zero of the real part) for all values of $\lambda_i$ in the domain $0\leq
\lambda_i \leq 1$.
This distortion, or some approximation to it, can be used in the argument
given above.

For the remaining integrations on the $dr_j$ $(j\leq i)$ one uses Propositions
4.4(ii) and 4.5, and Remark 4.3(ii).
This gives for the singular part of $F(q)$ at $\widehat{q}$ a function of form
F$'$, in some appropriately scaled
variable $\varphi(q)$, multiplied by an analytic function of $q$.

\newpage

\noindent{\bf 7. Computation for Self-Energy Case.}

\medskip

Contributions from photon lines $j$ coupled nonmeromorphically on both ends
into
the same side $s$ of the triangle were excluded from the discussion in
section 6.
For these values of $j$ one can, in order to exclude double counting, impose
the
condition $\lambda_j \geq \lambda'_j$, where $\lambda_j$ is the
$\lambda$-parameter associated with the nonmeromorphic coupling on the tail of
photon line $j$, and $\lambda'_j$ is the $\lambda$-parameter associated with
the
nonmeromorphic coupling on the head of line $j$.
Momentum $k_j$ flows along line $j$ from its tail to its head, according to our
conventions.

The formulas of section 2 of Ref. 1 refer to momentum $k_j$ flowing out of the
charged-particle line at the tail of the photon line $j$.
The coupling at the head can be treated like the coupling at the tail, but with
a reversal of the sign of $k_j$.
Then the effect of the two couplings into the same side $s$  is to replace
$p_s$
by $p_s + (\lambda_j - \lambda'_j) k_j$, and to integrate on
$\lambda_j$ from zero to one and on $\lambda'_j$ from zero to $\lambda_j$.
The condition Im $p_s\Omega_j >0$ then retains its usual from.
The reduction of the domain of integration does not upset the arguments of
section 6.

To bring this case into accord with section 6 we use the following
transformations:
$$
\eqalignno{
&\int^1_0 d\lambda \int^\lambda_0 d\lambda' f((\lambda -\lambda')k)\cr
=&\int^1_0 \lambda d\lambda \int^1_0 d\lambda'' f (\lambda(1-\lambda'')k)\cr
=&\int^1_0 \lambda d\lambda \int^1_0 d\lambda''' f(\lambda\lambda'''k)\cr
=&\int^1_0 dh f(hk) \int^1_0 \lambda d\lambda \int^1_0 d\lambda '''
\delta (h-\lambda\lambda''')\cr
=&\int^1_0 dh f(hk) \int^1_0 d\lambda \int^1_0 d\lambda'''
\delta ({h\over \lambda}-\lambda''')\cr
= &\int^1_0 dh f(hk) \int^1_h d\lambda\cr
= &\int^1_0 dh f(hk) (1-h).\cr}
$$
The variable $h$ plays the role played by $\lambda$ in section 6.

\newpage

{\bf References}

\noindent 1. T. Kawai and H.P. Stapp, {\it Quantum Electrodynamics at Large
Distances I:  Extracting the Correspondence--Principle Part.}
Lawrence Berkeley Laboratory Report LBL 35971. Submitted to Phys. Rev.\\
2. T. Kawai and H.P. Stapp {\it Quantum Electrodynamics at Large Distances II:
Nature of the Dominant Singularities.}
Lawrence Berkeley Laboratory Report LBL 35972. Submitted to Phys. Rev.\\
3. H.P. Stapp, {\it in } Structural Analysis of Collision Amplitudes, ed.
R. Balian and D. Iagolnitzer, North Holland, New York (1976)

\end{document}